\documentclass[english]{article}

\usepackage[T1]{fontenc}
\usepackage[latin9]{inputenc}
\usepackage{geometry}
\usepackage{amssymb}
\usepackage{graphicx}
\PassOptionsToPackage{normalem}{ulem}
\usepackage{ulem}

\makeatletter

\providecommand{\tabularnewline}{\\}

\makeatother

\usepackage{babel}
\begin{document}


\title{Equations of State and Maximum Mass of Neutron Stars in light of
PSR J1614-2230}

\author{Carlos Daniel Xu}
\maketitle
\begin{abstract}
We shall examine various types of equations of state for neutron stars,
which determine the structure of neutron stars. In particular, the
relation between mass and radius of neutron stars is of primary consideration.
By combining an equation of state (EOS) with the Tolmann-Oppenheimer-Volkoff
structure equations, we can determine the theoretical maximum mass
of a neutron star for a given equation of state. One question we seek
to answer is whether quark matter can exist in the core of a neutron
star. In light of the discovery of pulsar PSR J1614-2230, the mass
of which is observed to be $1.97\pm0.04M_{\odot}$ (solar masses),
a valid equation of state must achieve a maximum mass that is greater
than 2 solar masses. To try to solve this problem, we experiment with
different sets of parameters for the quark matter to try to meet the
lower limit 2-solar-mass criterion. It is found that certain parameters
contribute significantly to the maximum mass of a neutron star. 
\end{abstract}

\section{Introduction}
Neutron stars are fascinating objects of study for theoretical physics
because their densities are many times greater than that of matter
found on earth (imagine a Boeing 747 compressed to a grain of sand),
and rather little is known about their inner structure and their properties.
Due to this lack of knowledge, physicists have to speculate using
theoretical models. For a given equation of state (EOS), there are
structure equations known as Tolmann-Oppenheimer-Volkoff (TOV) Equations
which relate the mass and radius of a neutron star. The equation of
state is related to and depends upon the composition of the neutron
star. The problem at hand is to find an equation of state that, when
supplemented with the TOV equations, models the structure of a neutron
star. 

To solve this problem, we try different equations of state to model
neutron stars. Since the composition of neutron stars is not uniform,
multiple equations of state {}``stitched together'' are required
to maintain accuracy and validity. For example, a neutron star is
known to have two distinguishable layers of crust and a nuclear matter
core. It is generally accepted that a neutron star has an outer crust,
an inner crust, a nuclear matter core, and likely a quark matter core.
The composition of a neutron star is vital to modeling a neutron star,
and intuitively affects the equation of state. We first consider the
composition of the star, and their corresponding EOSs. Then, we calculate
the threshold pressures for the phase transitions between different
types of matter that make up a neutron star. Thus, we have a comprehensive
equation of state that is valid and defined throughout the star. Finally,
we solve the TOV equations using the comprehensive EOS to produce
a graph relating the mass and radius of a neutron star.  

One question we seek to answer is whether quark matter can exist in
the core of a neutron star, whilst achieving a maximum mass that is
greater than 2 solar masses in light of the discovery of pulsar PSR
J1614-2230, which has been observed to be $1.97\pm0.04M_{\odot}$.
Thus, the discovery of the pulsar sets a lower limit to the maximum
mass for a neutron star, as pictured in Figure 3 of the paper by Demorest
et al. {[}3{]}. Many theoretical equations of state can be ruled out
simply from this criterion. The central density of neutron stars is
known to be 5 to 10 times higher than the nuclear density, so baryonic
matter (such as neutrons and protons) should become quark matter.
However, previous research has shown that quark matter renders an
equation of state to be {}``softer'', ie. to have a lower maximum
mass. Therefore, it is interesting to see whether an EOS with quark
matter is still viable. In the quark matter equation of state there
are a few constants that can be manipulated to fit the 2-solar-mass
condition {[}1{]}. In this research it is shown that one of the manipulable
parameters more significantly affects the maximum mass value relative
to the other parameters.

\section{Tolmann-Oppenheimer-Volkoff Equations}

The Tolmann-Oppenheimer-Volkoff Equations are structure equations
used to relate the mass and radius of a neutron star. It is worth
noting that the TOV equations can be solved completely once we establish
$\rho(P)$, or the equation of state. One can consider a neutron star
as a fluid under gravitational pressure. In other words, the upper
layers of the star compress the lower layers; therefore, the pressure
and density are greatest at the center. For a Newtonian system, mathematically
one can establish that such system could be described by $\frac{dP(r)}{dr}=-\frac{GM(r)\rho(r)}{r^{2}}$.
In dwarf stars and other stars less massive than neutron stars, it
is usually permissible to ignore relativistic effects because they
are negligible and simply use the Newtonian relationship. For neutron
stars, however, it is necessary to make corrections; for example,
the intuitive equation $\frac{dP(r)}{dr}=-\frac{GM(r)\rho(r)}{r^{2}}$
must be corrected to become Equation 1. In astrophysics, the set of
equations to model the physical structure of spherically symmetric
stars in hydrostatic equilibrium is called the Tolmann-Oppenheimer-Volkoff
(TOV) equations. The set of TOV equations incorporate general relativistic
effects due to the immense density of a neutron star. We apply them
to find the mass and radius of a neutron star given an equation of
state (discussed later in this paper), which is dictated by the composition
of the neutron star. 

Derived from Einstein's equations of general relativity, the TOV equations
are

\begin{equation}
\frac{dP(r)}{dr}=-\frac{G}{r^{2}}\left[\rho(r)+\frac{P(r)}{c^{2}}\right]\left[M(r)+4\pi r^{3}\frac{P(r)}{c^{2}}\right]\left[1-\frac{2GM(r)}{c^{2}r}\right]^{-1}
\end{equation}

\begin{equation}
\frac{dM(r)}{dr}=4\pi\rho(r)r^{2}
\end{equation}

where $P(r)$ is the pressure at $r$, $\rho(r)$ is the mass density
at $r$, $M(r)$ is the mass enclosed by the sphere of radius $r$.
The former equation refers to the hydrostatic equilibrium and the
latter describes mass balance. These equations, when appended with
an equation of state (relating the pressure and density) $\rho(P)$,
can be solved for $P(r)$, $\rho(r)$, and $M(r)$, the last of which
provides information about the maximum mass, given an equation of
state. 

It is straightforward to note that inside the star, $P>0$ and the
pressure decreases monotonically as we move away from the center.
The boundary conditions needed to solve the TOV equations are $M(0)=0$,
$\rho(0)$ is the (unknown) central density, and $M(R)=$ mass of
neutron star, where $R$ is the radius of the neutron star. The central
density value affects the mass and corresponding radius of a neutron
star. By varying the central density, the maximum mass of neutron
stars can be found for a given equation of state.

\section{Equations of State }

In the previous section, we alluded to the missing component that
needs to supplement the Tolmann-Oppenheimer-Volkoff equations in order
to solve them and find the mass-radius relationship. An equation of
state is most usually in the functional form $\rho(P)$ relating the
pressure and the mass density. If for a given density $\rho$ an EOS
can sustain a larger pressure, then the EOS is considered {}``harder''
and produce a larger maximum mass. The opposite is true for a {}``softer''
EOS. Equations of state vary depending on the composition of a neutron
star. Thus, the equations of state required are distinct for different
distances $r$ from the center. For example, a neutron star's equation
of state will differ at the outer crust, inner crust, and at various
depths of the core. The most interesting EOS, which is unknown, is
at the densest part of the core, the composition of which we can only
speculate. The speculations on the composition of the core revolve
around meeting certain observational restraints, such as maximum mass
values. For example, an equation of state that produces a maximum
mass less than 2 solar masses is essentially ruled out.

\subsection{Outer Crust}

At the lowest densities of the neutron star, we find the outer crust.
It is mostly made up of atomic nuclei. The equation of state of the
outer crust is well-known and has been tabulated by Haensel and Pichon
(1994). Given the set of tabulated data $(P,\rho)$, we can use Mathematica
or similar computational software to interpolate the data points into
an approximate function for $\rho(P)$. The tabulated equation of
state for the outer crust is provided from pressures $3.833E24$ dyn
cm$^{-2}$ to $6.2149E29$ dyn cm$^{-2}$ in an EOS table from the
book by Haensel et al {[}7{]}. For pressures less than $3.833E24$
dyn cm$^{-2}$ we safely ignore those $(P,\rho)$ points because the
mass contribution relative to the overall mass is negligible.

\subsection{Inner Crust}

The next phase of the neutron star, at pressures above $6.2150E29$
dyn cm$^{-2}$, is the inner crust, composed of a plasma of nuclei
and electrons. The inner crust has a characteristic abundance of neutron-rich
nuclei in a {}``neutron fluid'' of free neutrons that have dripped
from nuclei. This neutron fluid differentiates the inner crust from
the outer crust. Roughly the same computational steps apply here,
where we take the tabulated results for $(P,\rho)$ from the SLy Inner
Crust EOS model seen in Table 3 of Douchin \& Haensel (2001) {[}4{]}
and interpolate using software to produce an equation of state. The
pressure values for this stage are roughly from $6.2150E29$ dyn cm$^{-2}$
to $5.3711E32$ dyn cm$^{-2}$.

\subsection{Neutron Matter Core}

Nuclei cannot exist at densities above $1.5E14$ g cm$^{-3}$ {[}7{]}
which corresponds to roughly to the pressure $6E32$ dyn cm$^{-2}$
; at such high densities, nuclei become a plasma of mostly neutrons
and some protons. In the inner crust, protons clustered to form nuclei.
In the core, it is energetically favourable for protons to not cluster,
hence the plasma of neutrons and protons instead of nuclei. The precise
calculation of the equation of state at this phase is a difficult
problem. The uncertainty in the calculations is larger in the core
than in the crust, simply because the matter in the core is at much
higher densities than anything observed.

For the neutron core, we use the EOS provided by Gandolfi et al {[}5{]}.
The results of their Quantum Monte Carlo method (QMC), a sophisticated
and accurate calculation, can be parametrized by $E$ the energy per
particle of neutron matter

\begin{equation}
E(n)=a\left(\frac{n}{n_{0}}\right)^{\alpha}+b\left(\frac{n}{n_{0}}\right)^{\beta}
\end{equation}

where $n$ is the (neutron) number density, coefficients $a$ and
$\alpha$ are sensitive for the low density range of the equation
of state, and $b$ and $\beta$ for higher densities. Here the nuclear
saturation density $n_{0}=0.16E39$ cm$^{-3}$. There are a few options
for the coefficients, based on different ways of calculating the three-body
3N interactive forces which contribute to the resulting Hamiltonian.
We shall use the parameters of Urbana IX (UIX): $a=13.4$ MeV, $b=5.62$
MeV, $\alpha=0.514$, $\beta=2.436$ {[}5{]}.

From the energy per particle $E$ we can calculate the chemical potential
per particle for the neutron core $\mu_{N}(n)$ 

\begin{equation}
\mu_{N}(n)=a(\alpha+1)\left(\frac{n}{n_{0}}\right)^{\alpha}+b(\beta+1)\left(\frac{n}{n_{0}}\right)^{\beta}
\end{equation}

and finally calculate the pertinent two equation of state quantities:
the neutron matter mass density $\rho_{N}$ and pressure $P_{N}$ 

\begin{equation}
\rho_{N}(n)=n(E(n)+m_{n}c^{2})/c^{2}
\end{equation}

\begin{equation}
P_{N}(n)=n(\mu(n)-E(n))
\end{equation}

where $m_{n}$ is the mass of a neutron and $c$ is the speed of light
in a vacuum {[}5{]}. The equation of state function $\rho(P)$ is
clearly parameterized by the variable $n$, and by evaluating both
Eq. 5 and Eq. 6 for a large range of values for $n$, say from $0.35n_{0}$
to $2n_{0}$ we can obtain the equation of state for the hadronic
part of the core.

\subsection{Quark Matter in the Core}

Lastly, at the densest part of the core, the physical composition
is still unknown. It has been predicted that at high enough density,
the nucleic and neutron matter becomes a quark-gluon plasma (QGP)
{[}2{]}. However, the exact transition point (threshold for phase
change) is not known. The EOS of quark matter has a different form
than that of neutron matter. Since we are working with QGP, we avoid
using number density as a parameter because there are no neutrons
in quark matter. Instead, we can use the chemical potential $\mu$
as the parameter. We define the energy density $\epsilon_{Q}$ for
the quark matter core as

\begin{equation}
\epsilon_{Q}(\mu)=\left(\frac{a_{4}n_{c}n_{f}\mu^{4}}{4\pi^{2}}-\frac{a_{2}n_{c}n_{f}\mu^{2}}{12\pi^{2}}+B\right)/(\hbar c)^{3}
\end{equation}

where $a_{4},a_{2}$ is a constant parameter and $n_{c},n_{f}=3$
are the number of colours and flavours of the quark matter, respectively
{[}1{]}. $B$ is known as the bag constant. Now, we define the quantities
$\rho$ and $P$ which form the basis for the quark matter equation
of state {[}1{]}:

\begin{equation}
\rho_{Q}(\mu)=\frac{\epsilon_{Q}(\mu)}{c^{2}}
\end{equation}

\begin{equation}
P_{Q}(\mu)=\left(\frac{a_{4}n_{c}n_{f}\mu^{4}}{12\pi^{2}}-\frac{a_{2}n_{c}n_{f}\mu^{2}}{12\pi^{2}}-B\right)/(\hbar c)^{3}
\end{equation}

Again, the equation of state function $\rho(P)$ can be obtained by
evaluating Equations 8 and 9 for various values of $\mu$. Recall
that there are three parameters $a_{4}$,$a_{2}$ and $B$ that we
have yet to define. Theorists essentially try to change these parameters
to fit observations such as the two-solar-mass maximum mass requirement
which is discussed in the next section. We can roughly estimate the
order of magnitude of these parameters, but the precise calculation
is difficult. Therefore, it is reasonable to experiment with a few
values within a reasonable range to understand the properties of quark
matter. A set of values for these parameters is also provided in the
next section to illustrate the effects of the parameters on the resulting
graphs. Thus, the EOS for the quark matter is an unsolved problem.

\section{Discussion: Structure of the Neutron Star and Maximum Mass}

Given all the equations of state $\rho(P)$ that cover all the values
of pressure possible within a neutron star, the theoretical neutron
star model is then constructible using the Tolmann-Oppenheimer-Volkoff
Equations. For pressure values between $3.833E24$ dyn cm$^{-2}$
and $5.3711E32$ dyn cm$^{-2}$, we easily use the well-defined equations
of state of the outer crust and inner crust. For the crust to core
phase transition, we use the condition that the pressure is the same
at the transition for the crust EOS and core EOS. The density of the
core at the crust-core interface is around $0.4n_{0}$. 

For the core transition between neutron matter (NM) to quark matter
(QM), we consider two possibilities. First that there is no transition
to quark matter (represented by dashed red line in the figures). For
the second case that there is a transition into a QM core, we need
to find the phase transition between NM and QM.

\subsection{Threshold Pressure Calculation at the Phase Transition}

The procedure for determining the boundary pressure between the NM
EOS and QM EOS is two-fold. Firstly, let us note that $\mu_{quark}=(\mu_{neutron}-938$
MeV)/3 where $\mu_{neutron}=\mu_{N}(n_{N})$ is the chemical potential
of the neutron core at the same boundary pressure value, for some
number density $n_{N}$. So the first condition is that the chemical
potential of the quark matter at the boundary pressure $\mu_{quark}$
must equal ($\mu_{N}(n_{N})$ - 938 MeV)/3. The rest mass of a neutron
is 938 MeV. Secondly, the pressure value calculated by Equation 6
$P_{N}(n_{N})$ must equal $P_{Q}(\mu_{quark})$ from Equation 9.

We now consider the coordinates $(\mu_{quark}$$,P_{Q}(\mu_{quark}))$
and $(\mu_{N}(n_{N})-$938 MeV)/3$,P_{N}(n_{N}))$. Since $n_{N}$
is defined to be such that $(\mu_{N}(n_{N})-$938 MeV)/3 $=\mu_{quark}$
(condition 1), and $P_{N}(n_{N})=P_{Q}(\mu_{quark})$ (condition 2),
the phase transition occurs where $(\mu_{quark}$$,P_{Q}(\mu_{quark}))$
is equal to $(\mu_{N}(n_{N})-938$ MeV)/3$,P_{N}(n_{N}))$. To find
such coinciding coordinates, we can graph $(\mu_{quark}$$,P_{Q}(\mu_{quark}))$
for various values of $\mu_{quark}$ and graph $(\mu_{N}(n_{N})-$938
MeV)/3$,P_{N}(n_{N}))$ for various values of $n_{N}$. Then, if $(\mu_{quark}$$,P_{Q}(\mu_{quark}))$
is equal to $(\mu_{N}(n_{N})-$938 MeV)/3$,P_{N}(n_{N}))$, then the
threshold pressure value between the neutron core and the quark core
is $P_{Q}(\mu_{quark})=P_{N}(n_{N})$. 

As we mentioned in the previous section, there are a few unknown parameters
which make the quark matter EOS uncertain. Recall that there were
three undefined parameters $a_{4}$,$a_{2}$ and $B$ for the quark
matter core. Let us try with the following definitions: 

\uline{Table 1}

\vspace{10bp}

\hspace{-0.5cm}%
\begin{tabular}{|c|c|c|c|c||c|}
\hline 
 & Colour on Graph & $a_{4}$ & $a_{2}$ & $B^{1/4}$ & Threshold Pressure $P$ in dyn cm$^{-2}$ (NM to QM)\tabularnewline
\hline 
\hline 
a) & Blue & 1 & 0 & 175 MeV & 2.110 E 34\tabularnewline
\hline 
b) & Green & 1 & (150 MeV$)^{2}$ & 175 MeV & 8.651 E 34\tabularnewline
\hline 
c) & Dark Yellow & 0.7 & (150 MeV$)^{2}$ & 141.2 MeV & 6.2478 E 33\tabularnewline
\hline 
d) & Black & 0.6 & (150 MeV$)^{2}$ & 141.2 MeV & 2.58919 E 35\tabularnewline
\hline 
e) & Dashed Red & n/a & n/a & n/a & n/a (no transition to quark)\tabularnewline
\hline 
\end{tabular}

\vspace{10bp}

When we implement these parameters, we fully define quark matter,
and can then find their corresponding boundary pressures between nuclear
matter and quark matter. The dashed red line represents the neutron
matter core without quark matter; hence, the parameters are not applicable.
This can be visualized using Figure 1. The intersections between dashed
red with the other 4 colours represent the boundary between the four
possible types of quark matter (QM) and the neutron matter (NM). \vspace{70pt}

\includegraphics[bb=-70bp 0bp -9bp 183bp,scale=0.8]{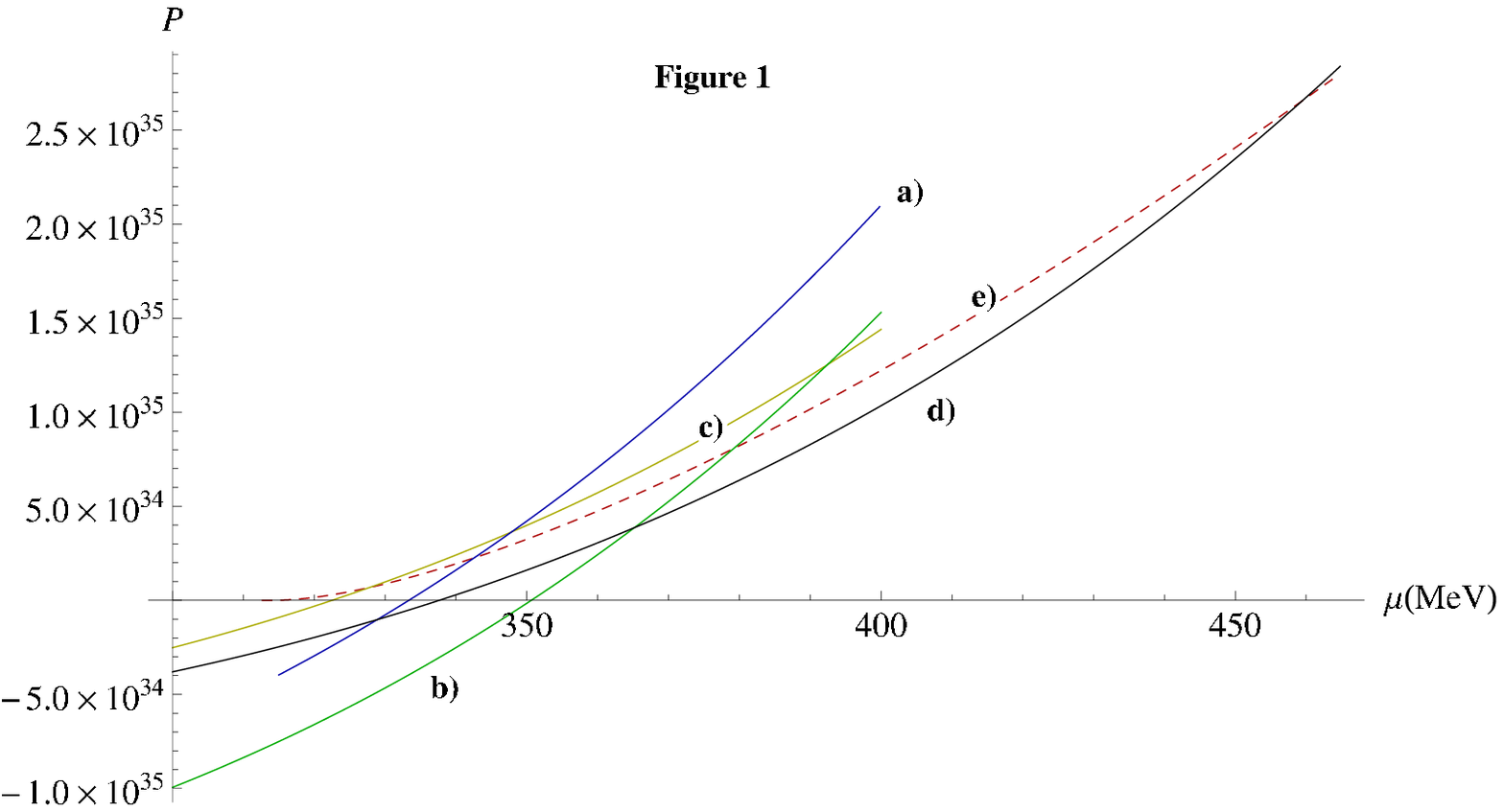}

\vspace{10bp}

Figure 1: The $P_{Q}(\mu_{quark})$ for the four theoretical types
of quark matter (blue, yellow, green, black) are plotted with that
of the neutron matter (dashed red). The x-axis values correspond to
the quark matter chemical potentials in MeV units. For the nuclear/neutron
matter (dashed red), we just have to use the relation $\mu_{quark}=(\mu_{neutron}-938$
MeV)/3, find $n_{N}$ such that $\mu_{N}(n_{N})=\mu_{neutron}$ and
evaluate with $P_{N}(n_{N})$ instead of $P_{Q}(\mu_{quark})$. Thus,
the intersections of the lines represent the threshold for phase transition
from NM to QM.

That is, at $P$ dyn cm$^{-2}$ and higher, the core is composed of
quark matter and the quark matter EOS is used. The following chart
summarizes the comprehensive equation of state that will be used in
the TOV equations. \vspace{10bp}

\hspace{70bp}%
\begin{tabular}{|c|c|}
\hline 
Equation of State & Pressure (g cm$^{-3}$)\tabularnewline
\hline 
Outer Crust & $3.833E24$ dyn cm$^{-2}$ to $6.2149E29$ dyn cm$^{-2}$ \tabularnewline
\hline 
Inner Crust & $6.2150E29$ dyn cm$^{-2}$ to $5.3711E32$ dyn cm$^{-2}$\tabularnewline
\hline 
Neutron Core & $5.3711E32$ dyn cm$^{-2}$ to $P$ (see chart above)\tabularnewline
\hline 
Quark Matter Core & $P$ to central density\tabularnewline
\hline 
\end{tabular}

\vspace{10bp}

\subsection{Computing the Maximum Mass}

Recall Equation 1 of the TOV equations. There are terms $\rho(r)$
that are unknown without an equation of state and prevent the solving
of the TOV equations. Now with a complete all-in-one equation of state,
as well as the threshold pressure for the phase transition between
QM and NM, we fully know $\rho(P)$ so we can write $\rho(r)$ as
$\rho(P(r))$. The TOV equation becomes 
\begin{equation}
\frac{dP(r)}{dr}=-\frac{G}{r^{2}}\left[\rho(P(r))+\frac{P(r)}{c^{2}}\right]\left[M(r)+4\pi r^{3}\frac{P(r)}{c^{2}}\right]\left[1-\frac{2GM(r)}{c^{2}r}\right]^{-1}
\end{equation}

\begin{equation}
\frac{dM(r)}{dr}=4\pi\rho(P(r))r^{2}
\end{equation}

The two functions $P(r)$ and $M(r)$ can be solved by Mathematica
easily, if we know $M(0)=0$ and $P(0)$, which is the central pressure.
Since the central pressure is unknown, and the value that we give
to $P(0)$ affects the solving of the functions $P(r)$ and $M(r)$,
we will get a distinct neutron star with a unique mass and radius
for every distinct value we assign to the central pressure. So if
we wanted to find the maximal mass possible for any neutron star given
our set of equations of state, we simply need to try a wide range
of central pressure values and see which one generates a neutron star
with highest mass. The graph generated from computing such maximum
mass is seen in Figure 2. This TOV-solving procedure is done for the
case in which we do not have quark matter in the core (dashed red)
, as well as for the 4 different types of quark matter. \vspace{1.5cm}

\includegraphics[bb=-40bp 0bp 373bp 256bp,scale=0.93]{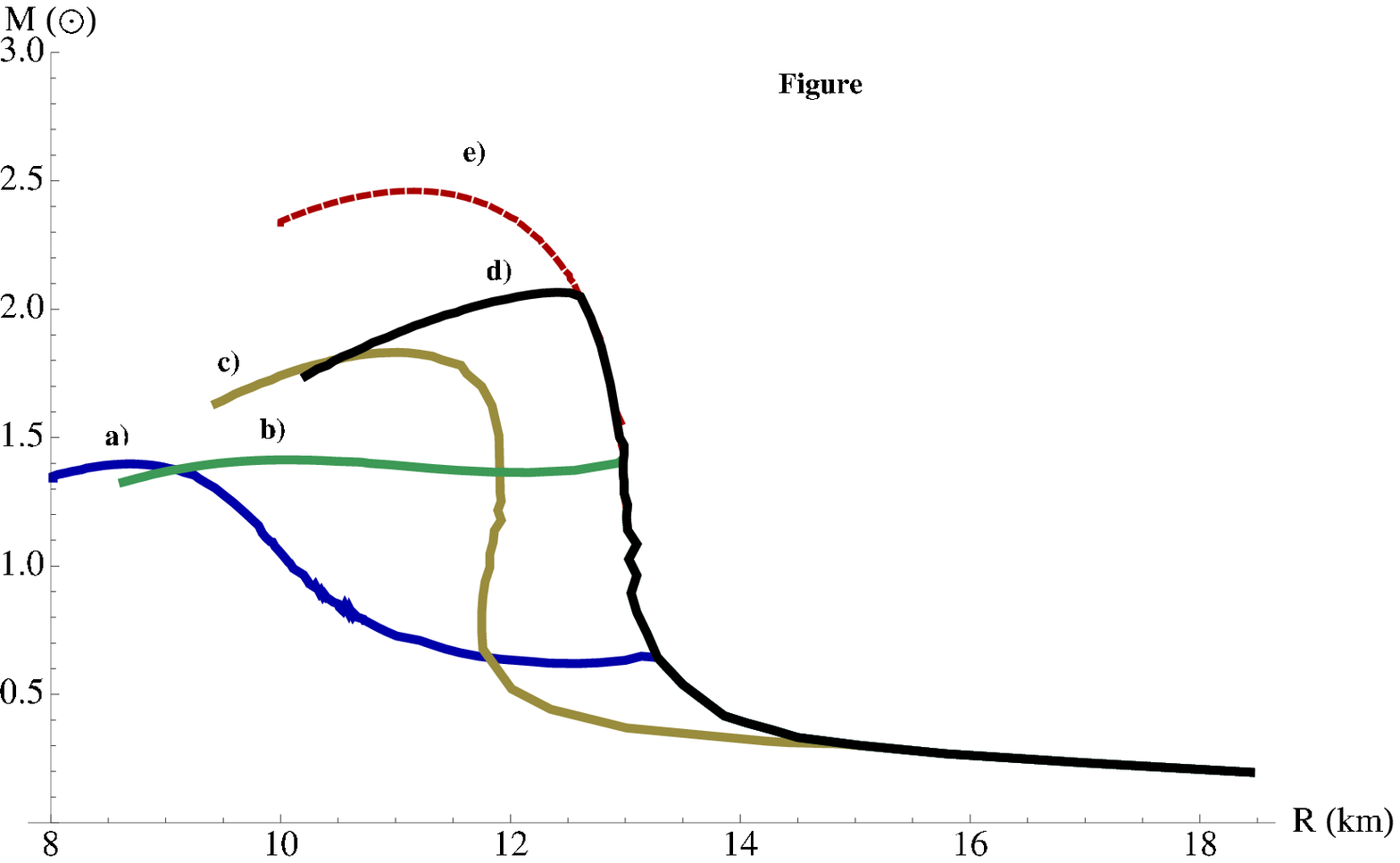}

Figure 2: The mass vs. radius graph shows the maximum mass possible
for the different equations of state. The highest valued maximum mass
belongs the equation of state without quark matter (dashed red), which
clearly shows that quark matter in the core {}``softens'' the overall
EOS. The next highest valued maximum mass corresponds to the equation
of state that has quark matter with the parameters of d) in Table
1 (black). In general, to interpret the Mass vs. Radius graphs such
as Figure 2, the central density value $\rho(0)$ used to solve the
TOV equations increases as you follow the curves from right to left.

\section{Conclusion: Two-Solar-Mass Neutron Star}

One of the more recent neutron star discoveries was that of the pulsar
PSR J1614-2230, the mass of which was determined to be $1.97\pm0.04M_{\odot}$
using Shapiro delay calculations by Demorest et. al {[}3{]}. This
revived interest in the research on the equations of state that met
the mass measurement of the PSR J1614-2230, and eliminated many possible
equation of state candidates. In other words, the discovery of the
two-solar-mass neutron star became a new benchmark for testing the
validity of a given equation of state. One of the prominent questions
that emerged was whether a hybrid equation of state with quark matter
like the one described in this paper would meet the two-solar-mass
criterion. The doubt arises from the fact that quark matter equation
of state is a {}``softer'' EOS, meaning it produces a smaller maximum
mass compared to any EOSs such as the neutron core EOS. 

As we can see from Figure 2 above, the hybrid equations of state with
the softer quark matter EOSs mostly produce maximum masses that do
not seem to meet the PSR J1614-2230 two-solar-mass criterion, whereas
the EOS without quark matter easily reaches 2.5$M_{\odot}$. The most
interesting hybrid EOS that still gives quark matter in the core is
the one given the parameters of d) in Table 1, which is represented
in Figure 1 and Figure 2 in black. This EOS with $a_{4}=0.6,a_{2}=$
(150 MeV$)^{2}$ and $B^{1/4}=141.2$ MeV produces a maximum mass
that safely meets the 2-solar-mass criterion. The other three quark
EOSs do not meet the 2-solar-mass criterion. Both the blue and green
Mass vs Radius curves (corresponding to parameters a) and b) respectively)
have a maximum mass of around 1.5$M_{\odot}$. The dark yellow curve,
which uses parameters c), is even closer at roughly 1.85$M_{\odot}$
These chosen values for the parameters were inspired by Alford et.
al in their paper {[}1{]}.

We can draw some strong conclusions simply from these five curves
representing four quark hybrid EOSs and one quark-free hybrid EOS.
The blue and green parameters suggest that the $a_{2}$ parameter
does not affect the maximum mass significantly. The blue curve uses
$a_{2}=0$, while the green curve uses $a_{2}=$ (150 MeV$)^{2}$.
Yet, there is no practical difference in the maximum mass between
the blue and the green curves. Parameter $a_{2}$ mainly affects where
NM transitions to QM in the core. The green curve has a much higher
threshold pressure for the phase transition from NM to QM: the blue
curve transitions at 2.11E34 dyn cm$^{-2}$ while the green curve
transitions at 8.651E34 dyn cm$^{-2}$. 

The main distinction between the dark yellow curve and the blue and
green curves is that the dark yellow curve has the parameter $a_{4}=0.7$,
while the blue and green curves have $a_{4}=1$. So this result strongly
suggests that the $a_{4}$ parameter is highly relevant in determining
the maximum mass of a neutron star with quark matter at its core.
To further support this conjecture, we consider the black curve, produced
with the set of parameters d). The black curve uses the same value
for $a_{2}$ and $B^{1/4}$ as the dark yellow curve. The only value
for which they differ is for $a_{4}$: black uses $a_{4}=0.6$ while
dark yellow uses $a_{4}=0.7$. This supports the thought that the
lower the $a_{4}$ parameter, the higher the maximum mass. 

So it would seem intuitive to simply use a relatively low $a_{4}$
parameter value and expect to get a hybrid EOS with quark matter to
produce a maximum mass value of at least 2-solar masses. However,
this intuitive hypothesis is not exactly true. When we try a lower
$a_{4}$ parameter, the $P_{Q}(\mu_{quark})$ curves seen in Figure
1 may not intersect reasonably, thus we would not have a defined phase
transition. With the knowledge that the $a_{4}$ parameter most significantly
affects the maximum mass value for a given EOS, it will be worth looking
into whether we can manipulate $a_{2}$ and $B$ such that the $P_{Q}(\mu_{quark})$
curves do intersect with the $P_{N}(n_{N})$ curve thus establishing
the phase transition threshold pressure. So far we have found one
set of parameters that meet the two-solar-mass criterion. With the
knowledge that the $a_{4}$ parameter is important, further research
may shed light on precisely calculating the parameters and finding
the corresponding equations of state.

\section*{Acknowledgements}

I thank Dr. Rishi Sharma from TRIUMF Laboratory for Particle and Nuclear
Physics for his mentorship, advice and infinite support in this research
project. I also thank the Canadian Young Scientist Journal for their
support by accepting this paper for publication in the February 2013
issue.

\end{document}